\documentclass[twocolumn]{svjour3}

\smartqed

\usepackage{graphicx}
\usepackage{amssymb}
\usepackage{epsf}
\usepackage{txfonts}
\usepackage{natbib}

\journalname{Nonlinear Dynamics}

\begin{document}

\title{Classifying orbits in the classical H\'{e}non-Heiles Hamiltonian system}

\author{Euaggelos E. Zotos}

\institute{Department of Physics, School of Science, \\
Aristotle University of Thessaloniki, \\
GR-541 24, Thessaloniki, Greece \\
Corresponding author's email: {evzotos@physics.auth.gr}
}

\date{Received: 1 May 2014 / Accepted: 21 October 2014 / Published online: 31 October 2014}

\titlerunning{Classifying orbits in the classical H\'{e}non-Heiles Hamiltonian system}

\authorrunning{E. E. Zotos}

\maketitle

\begin{abstract}
The H\'{e}non-Heiles potential is undoubtedly one of the most simple, classical and characteristic Hamiltonian systems. The aim of this work is to reveal the influence of the value of the total orbital energy, which is the only parameter of the system, on the different families of orbits, by monitoring how the percentage of chaotic orbits, as well as the percentages of orbits composing the main regular families evolve when energy varies. In particular, we conduct a thorough numerical investigation distinguishing between ordered and chaotic orbits, considering only bounded motion for several energy levels. The smaller alignment index (SALI) was computed by numerically integrating the equations of motion as well as the variational equations to extensive samples of orbits in order to distinguish safely between ordered and chaotic motion. In addition, a method based on the concept of spectral dynamics that utilizes the Fourier transform of the time series of each coordinate is used to identify the various families of regular orbits and also to recognize the secondary resonances that bifurcate from them. Our exploration takes place both in the physical $(x,y)$ and the phase $(y,\dot{y})$ space for a better understanding of the orbital properties of the system. It was found, that for low energy levels the motion is entirely regular being the box orbits the most populated family, while as the value of the energy increases chaos and several resonant families appear. We also observed, that the vast majority of the resonant orbits belong in fact in bifurcated families of the main 1:1 resonant family. We have also compared our results with previous similar outcomes obtained using different chaos indicators.


\end{abstract}

\section{Introduction}
\label{intro}

The interest in the existence of a third integral of motion for stars moving in the potential of a galaxy had been revived back in late 50s and in the early 60s. It was initially assumed that the potential had a symmetry and it was time-independent, therefore in cylindrical coordinates $(r,\theta,z)$ it would be only a function of $r$ and $z$. There should exist five integrals of motion that are constant for six-dimensional phase space. However, the integrals can be either isolating or non-isolating. The non-isolating integrals normally fill the entire available phase space and give no restriction to the orbit.

By the time when H\'{e}non and Heiles wrote their pioneer paper, there were only two known integrals of motion namely the total orbital energy and the angular momentum per unit mass of the star. It can easily be shown, that at least two of the integrals, in general, are non-isolating. It was also assumed, that the third integral was also non-isolating because no analytical solution had been found for it so far. Nevertheless, observations of stars near the Sun as well as numerical computations of orbits in some cases behaved as if they obeyed in three isolating integrals of motion.

H\'{e}non and Heiles set out to see if they could find any real proof of that there should exist a third isolating integral of motion. They did this by conducting numerical computations, even though they did not hold too hard to the astronomical meaning of the problem; they only demanded that the potential they investigated was axial symmetric. They also assumed that the motion was confined to a plane and went over to Cartesian phase space $(x,y,\dot{x},\dot{y})$. After some trials, they managed to come up with a valid potential. This potential is analytically simple so that the orbits could be calculated rather easily but it is still complicated enough so that the types of orbits are non trivial. This potential is now known as the H\'{e}non and Heiles potential [\citealp{HH64}].

Over the years, a huge load of research has been devoted on the H\'{e}non-Heiles system (e.g., [\citealp{W84}, \citealp{CTW82}, \citealp{F91}, \citealp{RGC93}, \citealp{dML99}, \citealp{AVS01}, \citealp{AS03}, \citealp{AVS03}, \citealp{B05}, \citealp{CMV05}, \citealp{SASL06}, \citealp{SSL07}, \citealp{BBS08}, \citealp{SS08}, \citealp{AVS09}, \citealp{BBS09}, \citealp{SHSL09}, \citealp{SS10}, \citealp{BSBS12}, \citealp{CSS13}]). At this point we should emphasize, that all the above-mentioned references regarding previous studies in the H\'{e}non-Heiles system are exemplary rather than exhaustive, taking into account that a vast quantity of related literature exists. The vast majority of these papers deals mostly either with the discrimination between regular and chaotic motion or with the escape properties of orbits. In the present paper, on the other hand, we proceed one step further contributing to this active field by classifying ordered orbits into different regular families and also monitoring how the percentages of the main families of orbits evolve as a function of the total orbital energy, considering only energy levels below the escape energy, where orbits are bounded and a variety of chaotic and several types of ordered motions exist. This is exactly what makes this paper novel.

In a thorough pioneer study [\citealp{LS92}], analyzed the orbital content in the coordinate planes of triaxial potentials but also in the meridional plane of axially symmetric model potentials, focusing on the regular families. Few years later, [\citealp{CA98}] developed a method based on the analysis of the Fourier spectrum of the orbits which can distinguish not only between regular and chaotic orbits, but also between loop, box, and other resonant orbits either in two or three dimensional potentials. This spectral method was improved and applied in [\citealp{MCW05}] in order to identify the different kinds of regular orbits in a self-consistent triaxial model. The same code was improved even further in [\citealp{ZC13}], when the influence of the central nucleus and of the isolated integrals of motion (angular momentum and energy) on the percentages of orbits in the meridional plane of an axisymmetric galactic model composed of a disk and a spherical nucleus were investigated.

The structure of the present paper is as follows: In Section \ref{galmod} we present a detailed description of the properties of the H\'{e}non-Heiles system. All the different computational methods used in order to determine the character as well as the classification of the orbits are described in Section \ref{compmeth}. In the following Section, we conduct a thorough analysis of several sets of initial conditions of orbits presenting in detail all the numerical results of our computations. Our article ends with Section \ref{disc}, where the discussion and the conclusions of this research are presented.

\section{Properties of the H\'{e}non-Heiles system}
\label{galmod}

One of the most paradigmatic model potentials for time-independent Hamiltonian systems with two degrees of freedom (2 d.o.f) is undoubtedly the H\'{e}non-Heiles system [\citealp{HH64}]. This particular system defines the motion of a test particle with unit mass and the corresponding two-dimensional potential is given by
\begin{equation}
V(x,y) = \frac{1}{2} \left(x^2 + y^2 \right) + x^2 y - \frac{1}{3}y^3.
\label{HHpot}
\end{equation}
It can be seen in Eq. (\ref{HHpot}) that the potential is in fact composed of two harmonic oscillators that has been coupled by the perturbation terms $x^2 y - 1/3 y^3$.

The H\'{e}non-Heiles potential along with the dihedral $D4$ potential [\citealp{AGK89}] and the Toda potential [\citealp{T67}] belong to a specialized category of potentials. These potentials were obviously not chosen with the aim of modeling any particular type of galaxy. Instead, in the pioneering spirit of H\'{e}non \& Heiles, the aim was to select potentials which are generic in their basic properties, but convenient computationally, so that large numbers of computations could be performed. Furthermore, the H\'{e}non-Heiles potential admits a discrete triangular rotation $D3$ $(2\pi/3)$ symmetry.

The basic equations of motion for a test particle with a unit mass are
\begin{eqnarray}
\ddot{x} &=& - \frac{\partial V}{\partial x} = - x - 2xy, \nonumber \\
\ddot{y} &=& - \frac{\partial V}{\partial y} = - x^2 - y + y^2,
\label{eqmot}
\end{eqnarray}
where, as usual, the dot indicates derivative with respect to the time. Furthermore, the variational equations governing the evolution of a deviation vector\footnote{If \textbf{$S$} is the $2N$ dimensional phase space (four dimensional, in our case) where the orbits of a dynamical system evolve on, then a deviation vector \vec{w}, which describes a small perturbation of a specific orbit \vec{x}, evolves on a $2N$ dimensional space \textbf{$T_{x}S$} tangent to \textbf{$S$}.} ${\bf{w}} \equiv (\delta x, \delta y, \delta \dot{x}, \delta \dot{y})$ are
\begin{eqnarray}
\dot{(\delta x)} &=& \delta \dot{x}, \nonumber \\
\dot{(\delta y)} &=& \delta \dot{y}, \nonumber \\
(\dot{\delta \dot{x}}) &=& -\frac{\partial^2 V}{\partial x^2}\delta x - \frac{\partial^2 V}{\partial x \partial y}\delta y = - \left(1 + 2y \right)\delta x - 2x \delta y, \nonumber \\
(\dot{\delta \dot{y}}) &=& -\frac{\partial^2 V}{\partial y \partial x}\delta x - \frac{\partial^2 V}{\partial y^2}\delta y = -2x \delta x - \left(1 - 2y \right)\delta y.
\label{vareq}
\end{eqnarray}

Consequently, the Hamiltonian to potential (\ref{HHpot}) reads
\begin{equation}
H = \frac{1}{2}\left(\dot{x}^2 + \dot{y}^2 + x^2 + y^2\right) + x^2 y - \frac{1}{3}y^3 = h,
\label{ham}
\end{equation}
where $\dot{x}$ and $\dot{y}$ are the momenta per unit mass conjugate to $x$ and $y$, respectively, while $h > 0$ is the numerical value of the Hamiltonian, which is conserved.

\begin{figure*}[!tH]
\centering
\resizebox{0.7\hsize}{!}{\includegraphics{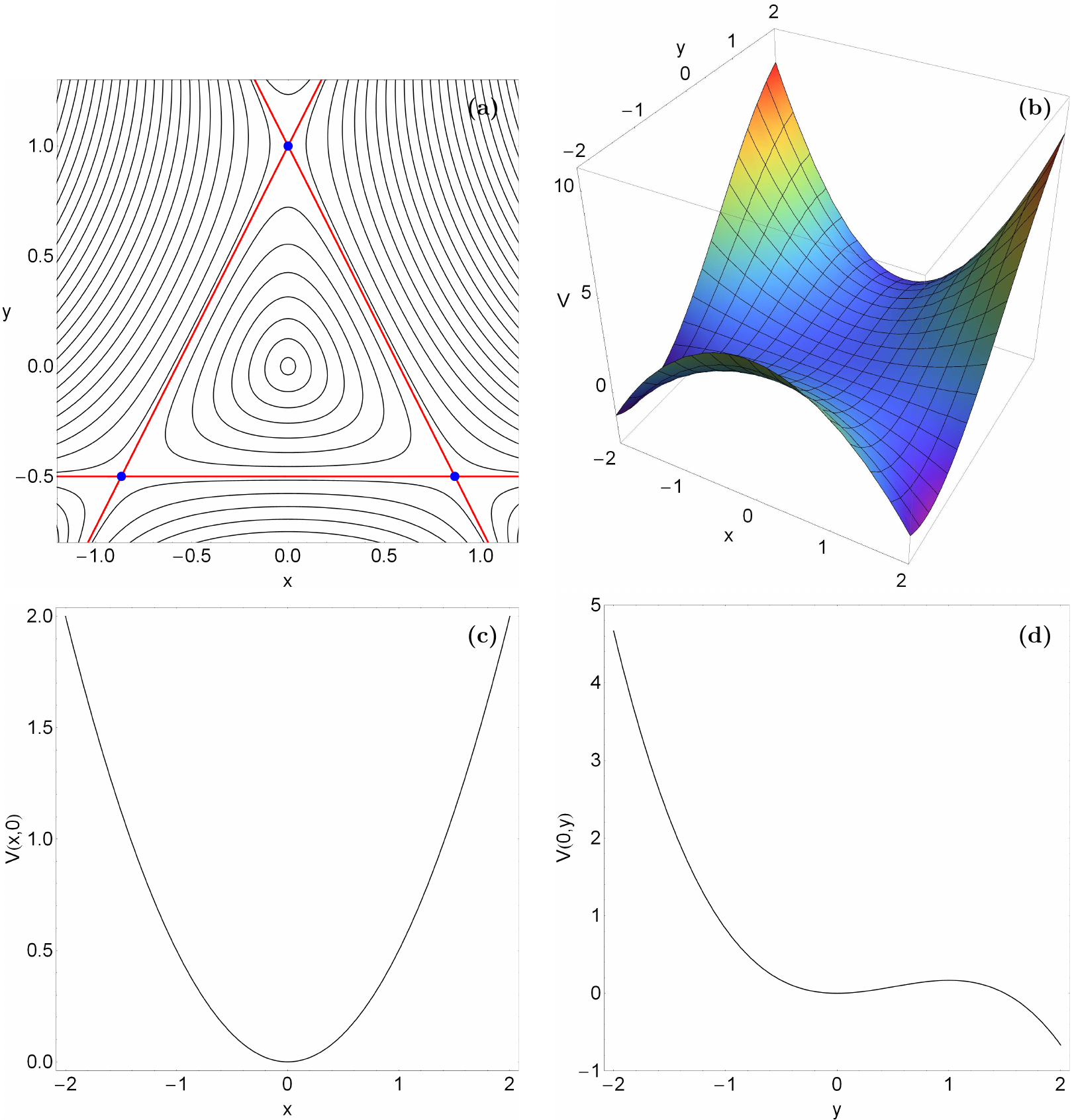}}
\caption{(a-upper left): Equipotential curves of the H\'{e}non-Heiles potential for various values of the energy $h$. The equipotential curve corresponding to the energy of escape is shown with red color, while the positions of the three saddle points are marked by blue dots; (b-upper right): The 3D surface of the potential $V(x,y)$; (c-lower left): The potential as a function of $x$ when $y = 0$; (d-lower right): The potential as a function of $y$ when $x = 0$.}
\label{HH}
\end{figure*}

The H\'{e}non-Heiles potential has a finite energy of escape $(h_{esc})$ which is equal to 1/6. For values of energy $h < h_{esc}$, the equipotential curves of the system are close thus making escape impossible. However, for larger energy levels $(h > h_{esc})$, the equipotential curves open and three exit channels appear through which the test particles may escape to infinity. The equipotential curves of the H\'{e}non-Heiles potential for various values of the energy $h$ are shown in Fig. \ref{HH}a. In the same plot, the equipotential corresponding to the energy of escape $h_{esc}$ is plotted with red color.

This potential has also some additional very interesting features. A three-dimensional surface of the potential $V(x,y)$ is presented in Fig. \ref{HH}b. For a constant value of $y$, $V(x,y)$ has the form of a parabola, $V(x,y) = (y + 0.5) x^2+ k(y)$ (the parabola when $y = 0$ is plotted in Fig. \ref{HH}c, while the potential as a function of $y$ when $x = 0$ is shown in Fig. \ref{HH}d). Furthermore, the potential has a stable equilibrium point at $(x,y) = (0,0)$ and three saddle points: $(x,y)$ = $\{(0,1)$, $(-\sqrt{3}/2 ,−1/2)$, $(\sqrt{3}/2 ,−1/2)\}$. The saddle points constitute the three corners of the equipotential curve $V(x,y) = 1/6$, that can be seen in Fig. \ref{HH}a. This triangular area that is bounded by the equipotential curve with energy $h = 1/6$ is the area of interest in our research.

\section{Description of the computational methods}
\label{compmeth}

In order to explore the orbital structure of the H\'{e}non-Heiles system, we need to define samples of orbits whose properties (regularity or chaos) will be identified. For this purpose, we define for each energy level (all tested energy levels are below the escape energy), dense grids of initial conditions regularly distributed in the area allowed by the value of the energy. Our investigation takes place both in the physical $(x,y)$ and the phase $(y,\dot{y})$ space for a better understanding of the orbital structure. In both cases, the step separation of the initial conditions along the $x$, $y$ and $y$ and $\dot{y}$ axes, respectively (in other words the density of the grids) was controlled in such a way that always there are at least 50000 orbits in every grid. For each initial condition, we integrated the equations of motion (\ref{eqmot}) as well as the variational equations (\ref{vareq}) using a double precision Bulirsch-Stoer FORTRAN algorithm (e.g., [\citealp{PTVF92}]) with a small time step of order of $10^{-2}$, which is sufficient enough for the desired accuracy of our computations (i.e., our results practically do not change by halving the time step). In all cases, the energy integral (Eq. (\ref{ham})) was conserved better than one part in $10^{-11}$, although for most orbits it was better than one part in $10^{-12}$.

Knowing whether an orbit is regular or chaotic is an issue of paramount importance. Over the years, several chaos indicators have been developed in order to determine the nature of orbits. In our investigation, we chose to use the Smaller ALingment Index (SALI) method [\citealp{S01}]. The SALI has been proved a very fast, reliable and effective tool, which is defined as
\begin{equation}
\rm SALI(t) = min\{d_- , d_+ \},
\label{sali}
\end{equation}
where $d_- \equiv \| {\bf{w_1}}(t) - {\bf{w_2}}(t) \|$ and $d_+ \equiv \| {\bf{w_1}}(t) + {\bf{w_2}}(t) \|$ are the alignments indices, while ${\bf{w_1}}(t)$ and ${\bf{w_2}}(t)$, are two deviation vectors which initially point in two random directions. For distinguishing between ordered and chaotic motion, all we have to do is to compute the SALI for a relatively short time interval of numerical integration $t_{max}$. In particular, we track simultaneously the time-evolution of the main orbit itself as well as the two deviation vectors ${\bf{w_1}}(t)$ and ${\bf{w_2}}(t)$ in order to compute the SALI. The variational equations (\ref{vareq}), as usual, are used for the evolution and computation of the deviation vectors. We should point out that the main feature of SALI which essentially distinguishes it from the computation of the maximal Lyapunov characteristic exponent (mLCE) is the introduction of one additional deviation vector with respect to a reference orbit. Indeed, by considering the relation between two deviation vectors (instead of one deviation vector and the reference orbit in the case of the mLCE), one is able to circumvent the difficulty of the slow convergence of mLCE to zero (or non-zero) values as time tends to infinity. In other word, SALI is computationally faster than mLCE, which requires longer integration times in order to it reveal the true chaotic nature of an orbit (see e.g., [\citealp{ZCar14}]).

\begin{figure}[!tH]
\centering
\includegraphics[width=\hsize]{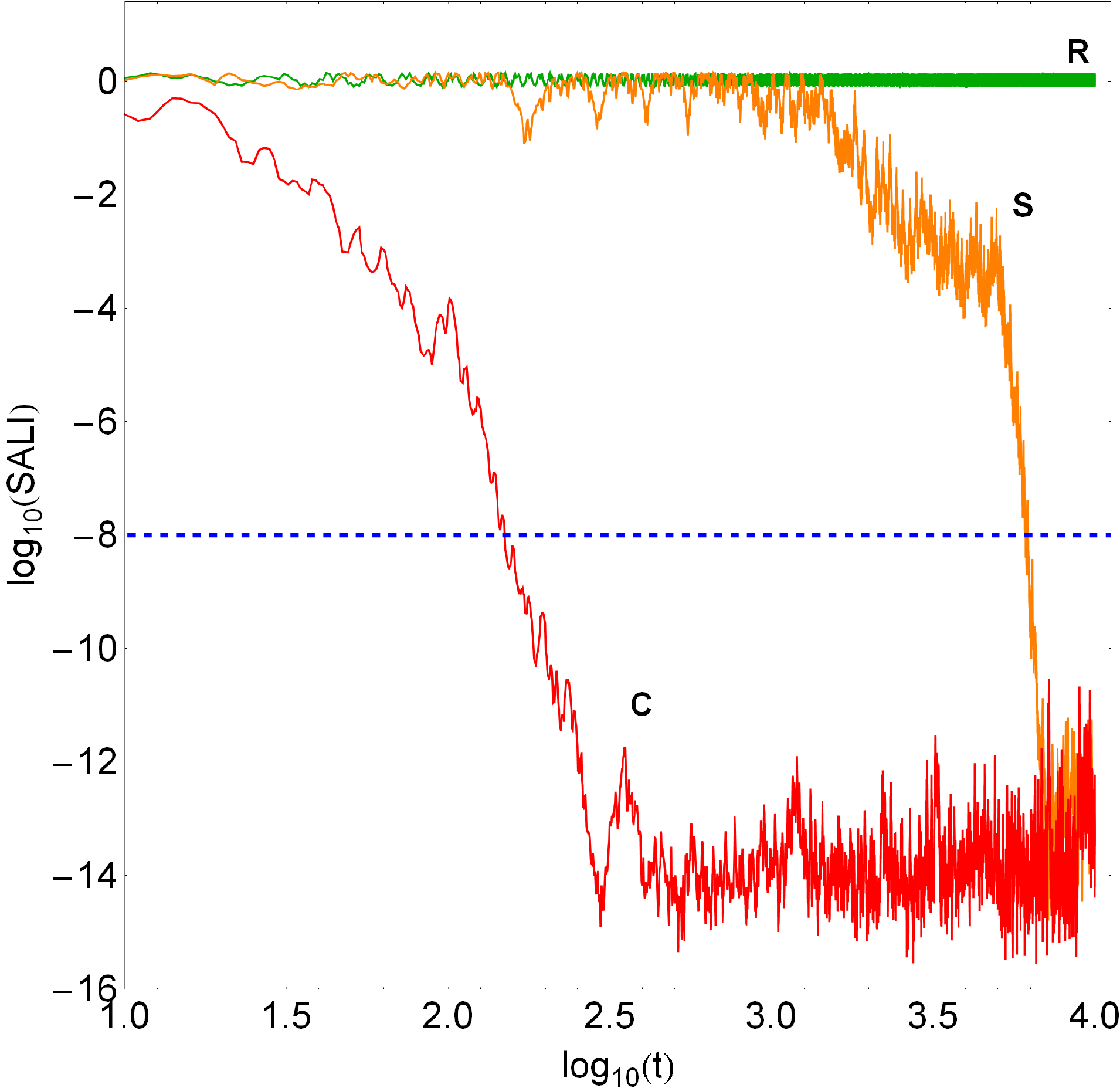}
\caption{Time-evolution of the SALI of a regular orbit (green color - R), a sticky orbit (orange color - S) and a chaotic orbit (red color - C) in the H\'{e}non-Heiles system for a time period of $10^4$ time units. The horizontal, blue, dashed line corresponds to the threshold value $10^{-8}$ which separates regular from chaotic motion. The chaotic orbit needs only about 150 time units in order to cross the threshold, while on the other hand, the sticky orbit requires a considerable longer integration time of about 6100 time units so as to reveal its true chaotic nature.}
\label{SALIevol}
\end{figure}

\begin{figure}[!tH]
\centering
\includegraphics[width=\hsize]{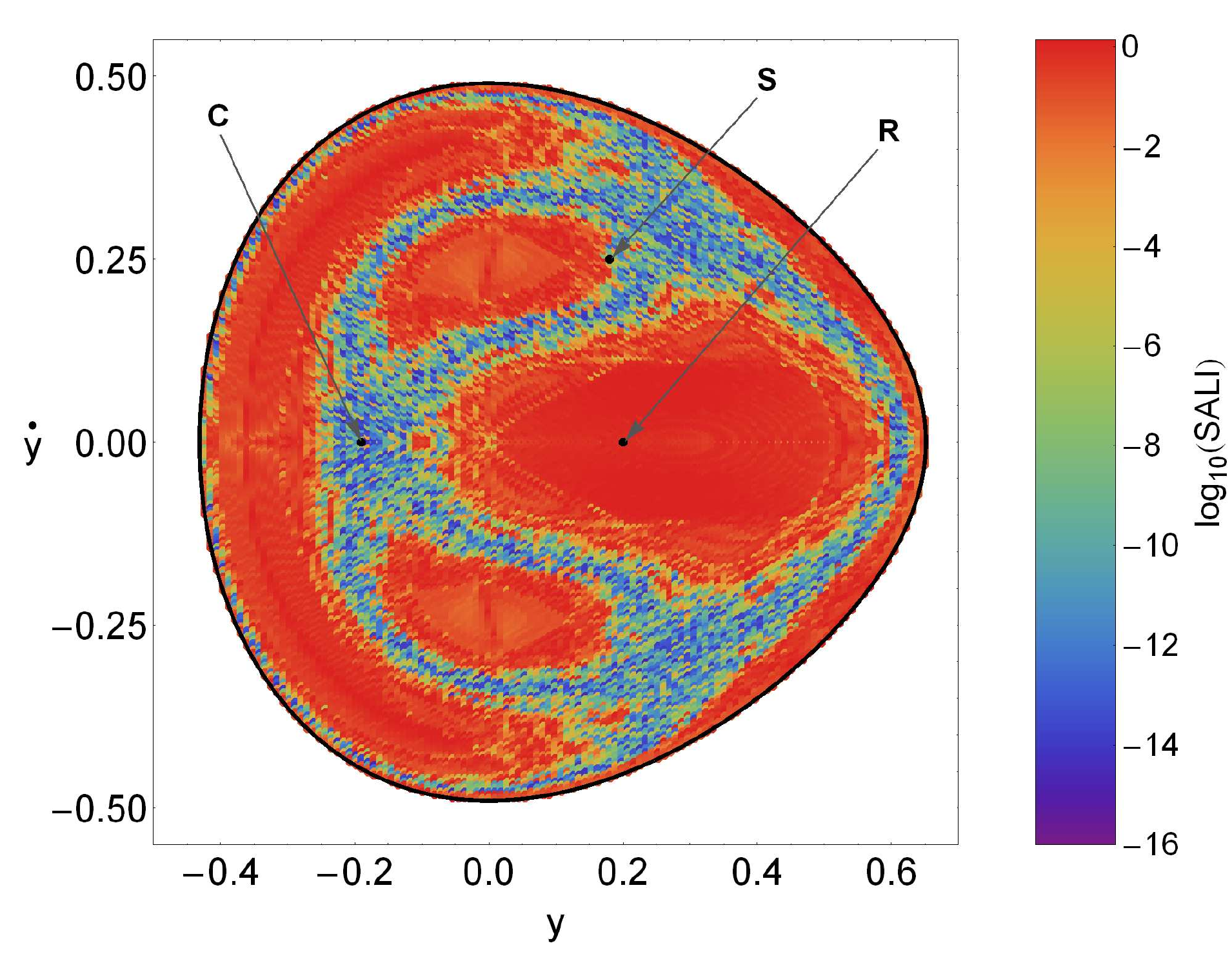}
\caption{Regions of different values of the SALI on the $(y,\dot{y})$ phase plane when the orbital energy is $h = 0.12$. Light reddish colors correspond to ordered motion, dark blue/purplr colors indicate chaotic motion, while all intermediate colors suggest sticky orbits. The three black dots point the initial conditions of the orbits of Fig. \ref{SALIevol}.}
\label{SALIgrid}
\end{figure}

The time-evolution of SALI strongly depends on the nature of the computed orbit since when the orbit is regular the SALI exhibits small fluctuations around non zero values, while on the other hand, in the case of chaotic orbits the SALI after a small transient period it tends exponentially to zero approaching the limit of the accuracy of the computer $(10^{-16})$. Therefore, the particular time-evolution of the SALI allow us to distinguish fast and safely between regular and chaotic motion. The time-evolution of a regular (R) and a chaotic (C) orbit for a time period of $10^4$ time units is presented in Fig. \ref{SALIevol}. We observe, that both regular and chaotic orbits exhibit the expected behavior. Nevertheless, we have to define a specific numerical threshold value for determining the transition from regularity to chaos. After conducting extensive numerical experiments, integrating many sets of orbits, we conclude that a safe threshold value for the SALI taking into account the total integration time of $10^4$ time units is the value $10^{-8}$. The horizontal, blue, dashed line in Fig. \ref{SALIevol} corresponds to that threshold value which separates regular from chaotic motion. In order to decide whether an orbit is regular or chaotic, one may use the usual method according to which we check after a certain and predefined time interval of numerical integration, if the value of SALI has become less than the established threshold value. Therefore, if SALI $\leq 10^{-8}$ the orbit is chaotic, while if SALI $ > 10^{-8}$ the orbit is regular. In Fig. \ref{SALIgrid} we present a dense grid of initial conditions $(y_0,\dot{y_0})$, where the values of the SALI are plotted using different colors. We clearly observe several regions of regularity indicated by light reddish colors as well as a unified chaotic domain (blue/purple dots). All intermediate colors correspond to sticky orbits. The initial conditions of the three orbits (regular, sticky and chaotic) whose SALI time-evolution was given in Fig. \ref{SALIevol}, are pinpointed with black dots in Fig. \ref{SALIgrid}. Therefore, the distinction between regular and chaotic motion is clear and beyond any doubt when using the SALI method.

In our study, each orbit was integrated numerically for a time interval of $10^4$ time units (10 billion yr), which corresponds to a time span of the order of hundreds of orbital periods. The particular choice of the total integration time is an element of great importance, especially in the case of the so called ``sticky orbits" (i.e., chaotic orbits that behave as regular ones during long periods of time). A sticky orbit could be easily misclassified as regular by any chaos indicator\footnote{Generally, dynamical methods are broadly split into two types: (i) those based on the evolution of sets of deviation vectors in order to characterize an orbit and (ii) those based on the frequencies of the orbits which extract information about the nature of motion only through the basic orbital elements without the use of deviation vectors.}, if the total integration interval is too small, so that the orbit do not have enough time in order to reveal its true chaotic character. Thus, all the sets of orbits of a given grid were integrated, as we already said, for $10^4$ time units, thus avoiding sticky orbits with a stickiness at least of the order of a Hubble time. All the sticky orbits which do not show any signs of chaoticity for $10^4$ time units are counted as regular ones, since that vast sticky periods are completely out of scope of our research. A characteristic example of a sticky orbit (S) in our galactic model can be seen in Fig. \ref{SALIevol}, where we observe that the chaotic character of the particular sticky orbit is revealed only after a considerable long integration time of about 6100 time units.

A first step towards the understanding of the overall orbital behavior of our galactic system is knowing the regular or chaotic nature of orbits. In addition, of particular interest is the distribution of regular orbits into different families. Therefore, once the orbits have been characterized as regular or chaotic, we then further classified the regular orbits into different families, by using the frequency analysis of [\citealp{CA98}]. Initially, [\citealp{BS82}, \citealp{BS84}] proposed a technique, dubbed spectral dynamics, for this particular purpose. Later on, this method has been extended and improved by [\citealp{SN96}] and [\citealp{CA98}]. In a recent work, [\citealp{ZC13}] the algorithm was refined even further so it can be used to classify orbits in the meridional plane. In general terms, this method calculates the Fourier transform of the coordinates of an orbit, identifies its peaks, extracts the corresponding frequencies and search for the fundamental frequencies and their possible resonances. Thus, we can easily identify the various families of regular orbits and also recognize the secondary resonances that bifurcate from them. The very same algorithm was used in several previous related papers (e.g., [\citealp{ZC13}, \citealp{CZ13}, \citealp{ZCar13}, \citealp{ZCar14}]).

\section{Numerical results \& Orbit classification}
\label{orbclas}

\begin{figure*}[!tH]
\centering
\resizebox{0.9\hsize}{!}{\includegraphics{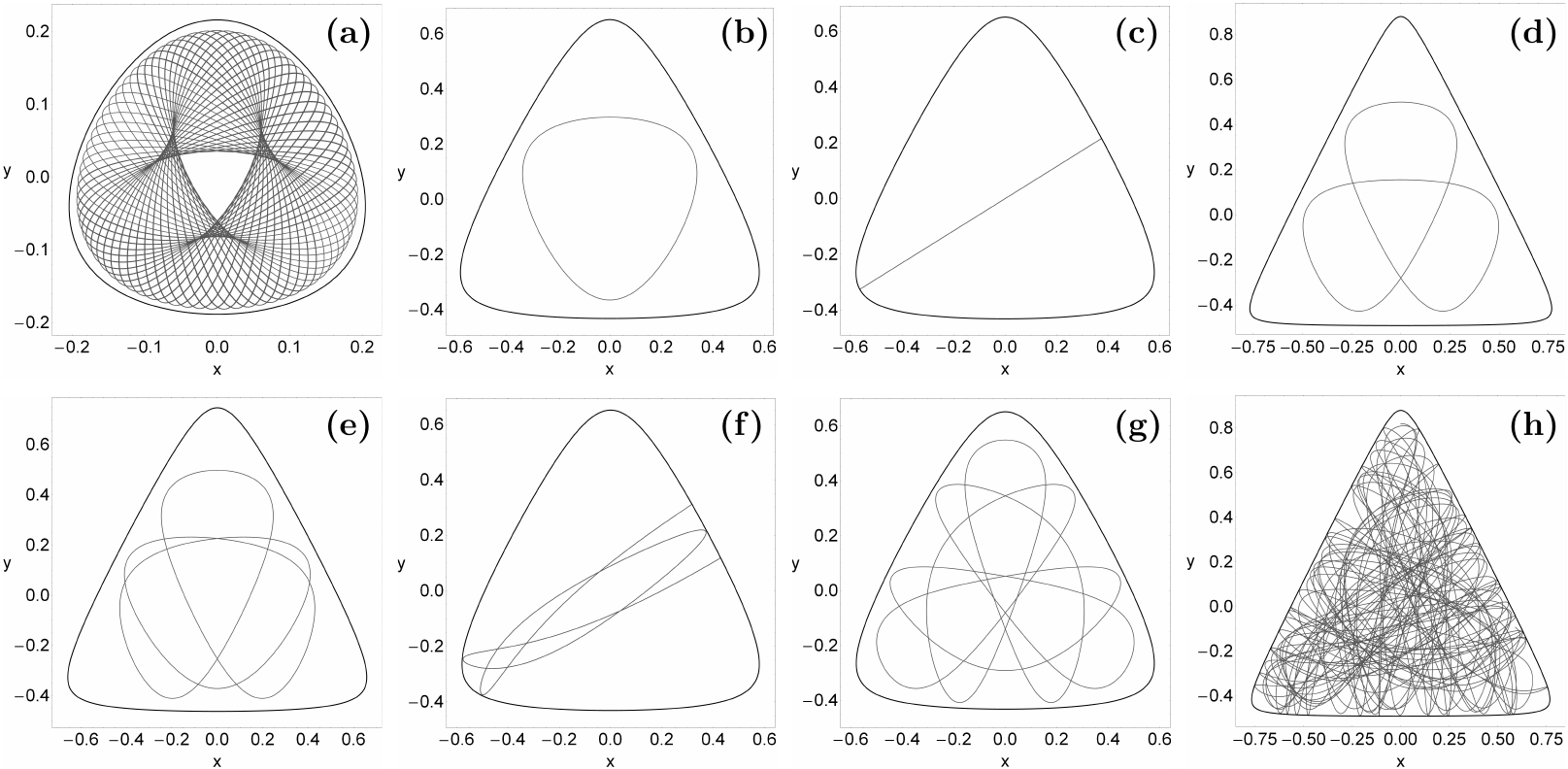}}
\caption{Collection of the basic types of orbits encountered in the H\'{e}non-Heiles system: (a) box orbit; (b) 1:1 resonant loop orbit; (c) 1:1 resonant linear orbit; (d) 2:2 resonant orbit; (e) 3:3 resonant orbit (f) 4:4 resonant orbit; (g) 5:5 resonant orbit; (h) chaotic orbit.}
\label{orbs}
\end{figure*}

\begin{figure*}[!tH]
\centering
\resizebox{0.9\hsize}{!}{\includegraphics{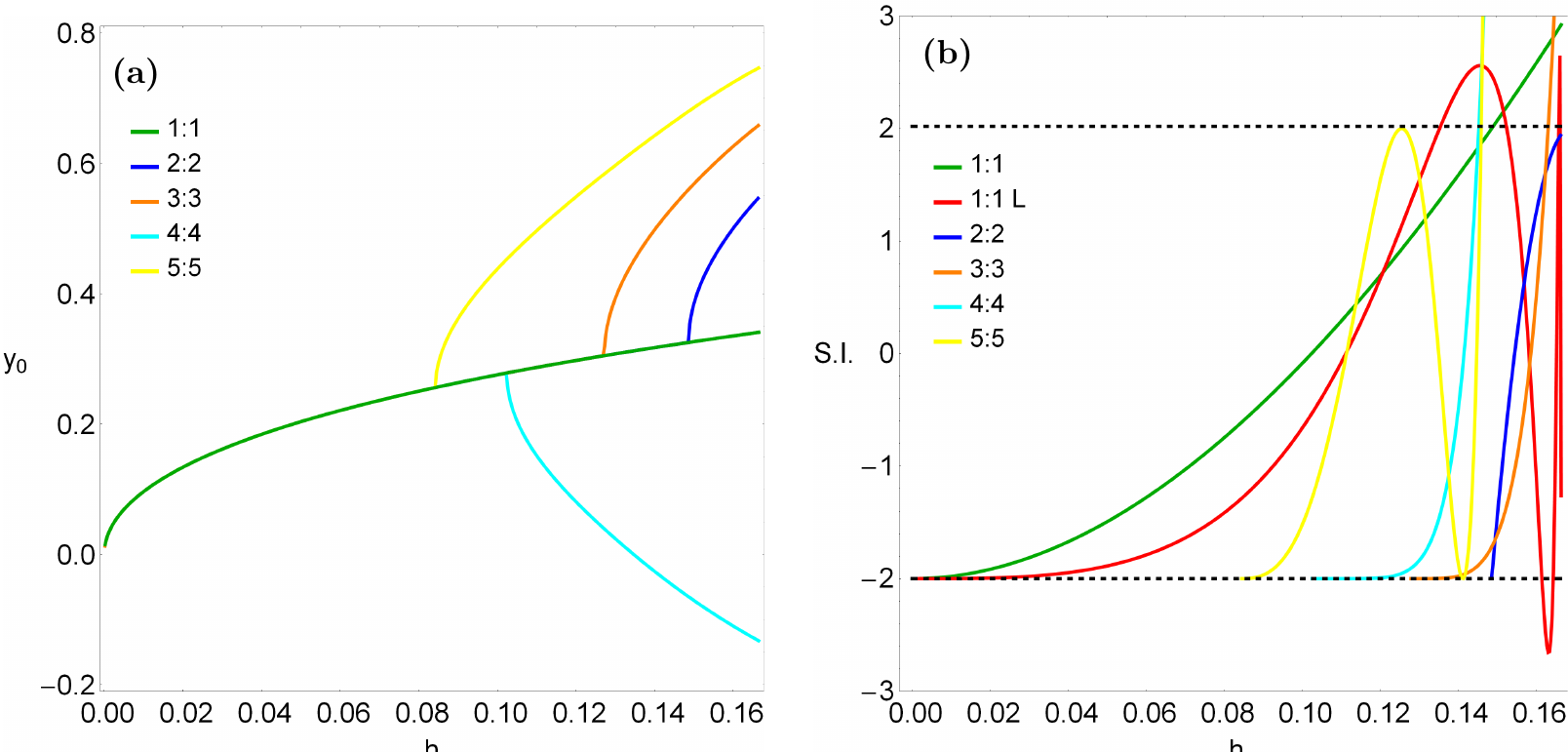}}
\caption{(a-left): The $(y_0,h)$ characteristic curves of the orbital families and (b-right): Evolution of the stability index S.I. of the families of periodic orbits shown in Figs. \ref{otd}a. The black dashed horizontal lines at -2 and +2 delimit the range of S.I. for which the periodic orbits are stable. The color code is the same in all panels.}
\label{otd}
\end{figure*}

\begin{figure*}[!tH]
\centering
\resizebox{0.9\hsize}{!}{\includegraphics{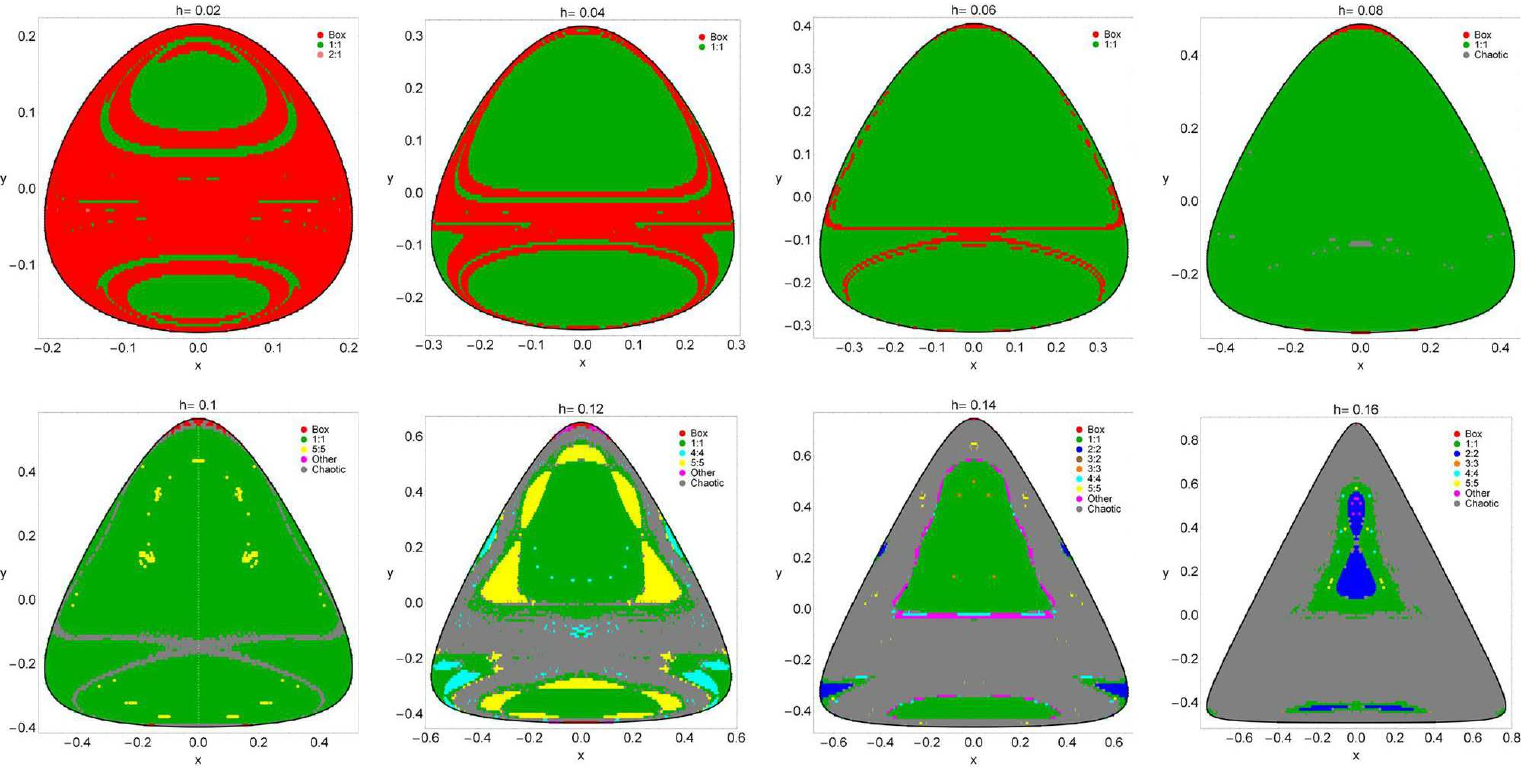}}
\caption{The orbital structure of the physical $(x,y)$ plane for several values of the energy $h$, distinguishing between chaotic and different types of regular orbits.}
\label{grd1}
\end{figure*}

\begin{figure*}[!tH]
\centering
\resizebox{0.9\hsize}{!}{\includegraphics{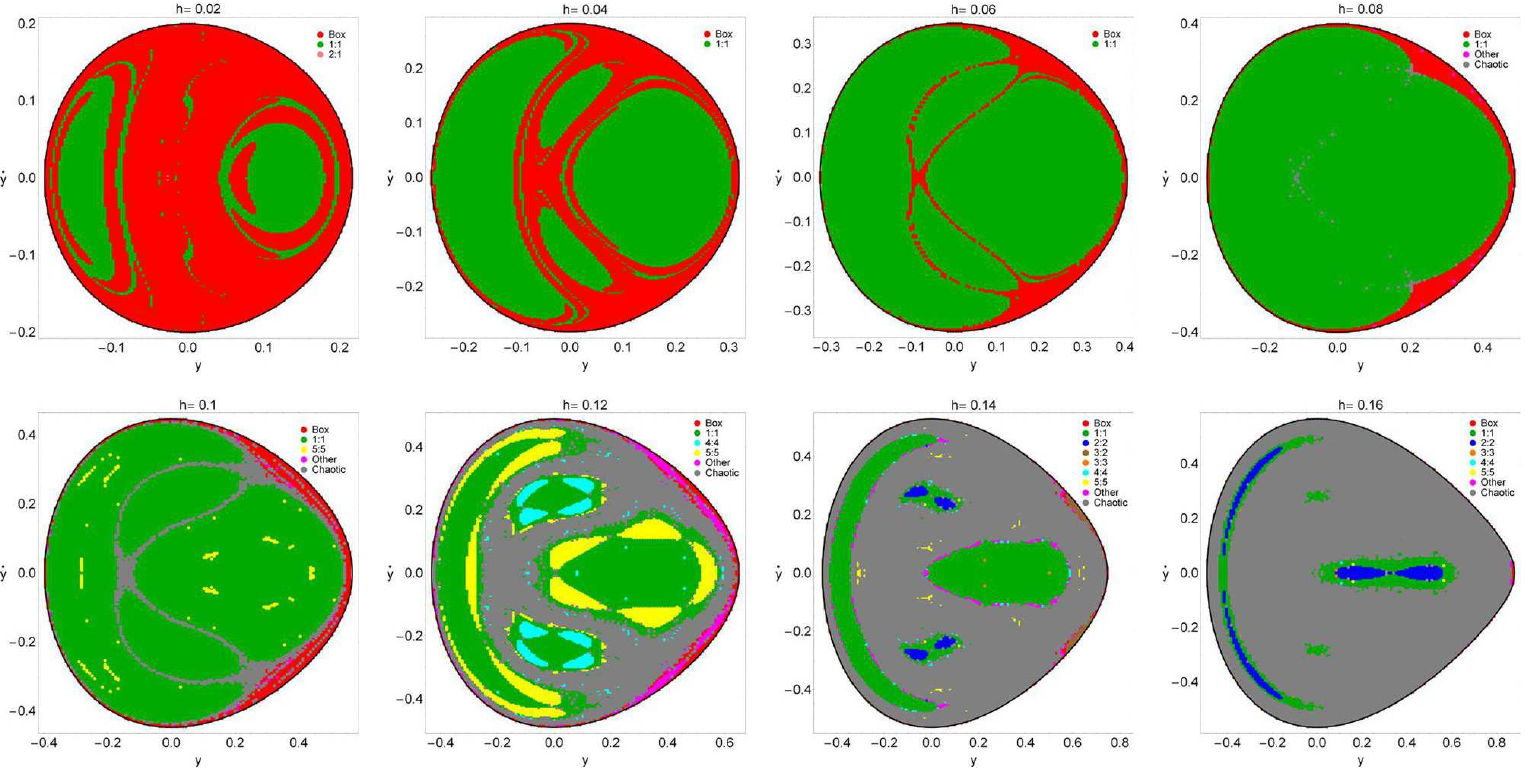}}
\caption{The orbital structure of the phase $(y,\dot{y})$ plane for several values of the energy $h$, distinguishing between chaotic and different types of regular orbits.}
\label{grd2}
\end{figure*}

\begin{figure*}[!tH]
\centering
\resizebox{0.9\hsize}{!}{\includegraphics{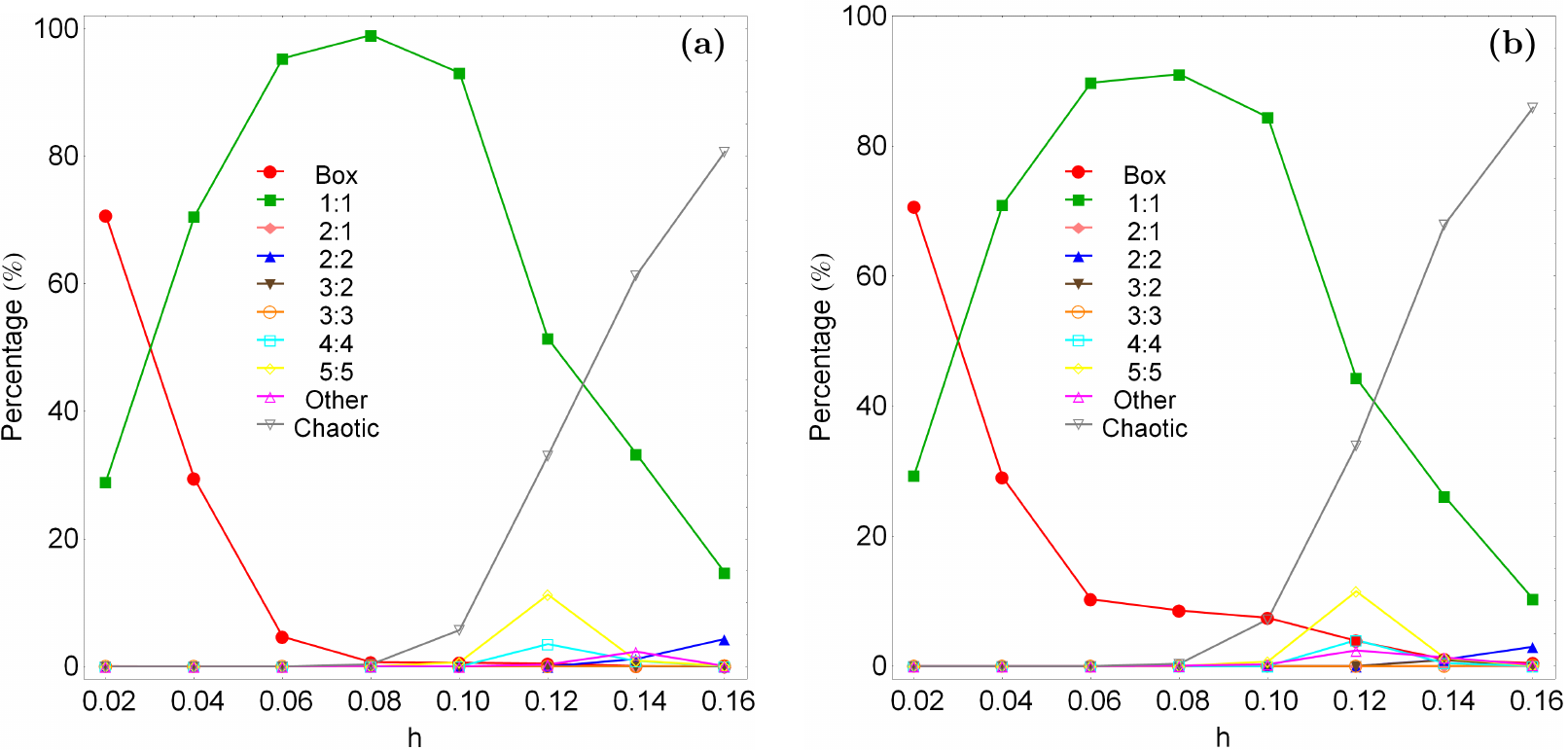}}
\caption{Evolution of the percentages of the different types of orbits when varying the energy $h$ (a-left): on the physical $(x,y)$ plane and (b-right): on the phase $(y,\dot{y})$ plane.}
\label{percs}
\end{figure*}

\begin{table}
\begin{center}
   \centering
   \caption{Type, initial conditions and the value of the energy of the orbits shown in Fig. \ref{orbs}(a-h). In all cases, $x_0 = 0$, $\dot{x_0}$ is found from the energy integral (Eq. (\ref{ham})), while $T_{\rm per}$ is the period of the resonant parent periodic orbits.}
   \label{table}
   \setlength{\tabcolsep}{2.0pt}
   \begin{tabular}{@{}lccccr}
      \hline
      Figure & Type & $h$ & $y_0$ & $\dot{y_0}$ & $T_{\rm per}$  \\
      \hline
      \ref{orbs}a & box        & 0.02 & -0.08200000 & 0.00000000 &           - \\
      \ref{orbs}b & 1:1 loop   & 0.12 &  0.29754452 & 0.00000000 &  6.08239553 \\
      \ref{orbs}c & 1:1 linear & 0.12 &  0.00000000 & 0.24494897 &  7.43702045 \\
      \ref{orbs}d & 2:2        & 0.16 &  0.50003040 & 0.00000000 & 12.47212865 \\
      \ref{orbs}e & 3:3        & 0.14 &  0.49779098 & 0.00000000 & 18.55933975 \\
      \ref{orbs}f & 4:4        & 0.12 &  0.10124417 & 0.28443422 & 29.16285964 \\
      \ref{orbs}g & 5:5        & 0.12 &  0.54846395 & 0.00000000 & 32.13261997 \\
      \ref{orbs}h & chaotic    & 0.16 &  0.82000000 & 0.00000000 &           - \\
      \hline
   \end{tabular}
\end{center}
\end{table}

The main objective of this numerical investigation is to distinguish between ordered and chaotic orbits for values of energy always lower than the escape energy. In order to accomplish a spherical view of the orbital properties of the Hamiltonian system, we expand our investigation in both the physical and the phase space. In both cases, the grids of initial conditions of orbits whose properties will be examined are defined as follows: for the physical $(x,y)$ space we consider orbits with initial conditions $(x_0, y_0)$ with $\dot{y_0} = 0$, while the initial value of $\dot{x_0}$ is obtained from the energy integral (\ref{ham}) as $\dot{x_0} = \dot{x}(x_0,y_0,h) > 0$. Similarly, for the phase $(y,\dot{y})$ space we consider orbits with initial conditions $(y_0, \dot{y_0})$ with $x_0 = 0$, while again the initial value of $\dot{x_0}$ is obtained from the energy integral (\ref{ham}). As we explained earlier, in our investigation we shall deal only with bounded motion $(h \in (0,h_{esc}))$ of test particles for values of energy in the set $h = \{0.02, 0.04, 0.06, 0.08, 0.10, 0.12, 0.14, 0.16\}$. The case of unbounded motion $(h > h_{esc})$ in the H\'{e}non-Heiles system has been investigated extensively in many previous papers (e.g., [\citealp{AVS01}, \citealp{AS03}, \citealp{AVS03}, \citealp{SASL06}, \citealp{SSL07}, \citealp{BBS08}, \citealp{SS08}, \citealp{AVS09}, \citealp{BBS09}, \citealp{SHSL09}, \citealp{SS10}, \citealp{BSBS12}]).

At this point, we would like to clarify some issues regarding the nomenclature of the orbits in our model. In the present paper, we decided to follow the classification method according to which, the orbits are separated into three main categories: (i) box orbits, (ii) $n:m$ resonant orbits, and (iii) chaotic orbits. According to our notation, all resonant orbits have the following recognizable $n:m$ oscillatory pattern: a resonant orbit completes $m$ oscillations perpendicular to the $x$ axis in the time that it takes the orbit to perform $n$ circuits along the $y$ axis. Furthermore, a $n:m$ resonant orbit would be represented by $n$ distinct islands of invariant curves in the $(x,\dot{x})$ phase plane and $m$ distinct islands of invariant curves in the $(y,\dot{y})$ surface of section (e.g., [\citealp{BT08}]). The $n:m$ notation we use for the regular orbits is according to [\citealp{CA98}, \citealp{ZC13}], where the ratio of those integers corresponds to the ratio of the main frequencies of the orbit, where main frequency is the frequency of greatest amplitude in each coordinate. Main amplitudes, when having a rational ratio, define the resonances of an orbit. In our research, we searched for resonant orbits $n:m$ up to $n,m \leq 10$ therefore, for all higher resonant orbits (if any) the numerical code assigns ``box" classification (this is a usual technique in orbit classification), which is indeed correct for $n \neq m$ (high resonant box orbits (e.g., [\citealp{CB82}]).

Our numerical calculations suggest that in the H\'{e}non-Heiles system apart from the usual box and chaotic orbits there are also many types of resonant periodic orbits. The majority of resonant orbits belong to the main 1:1 resonant family however, several secondary bifurcated families (i.e., the 2:2, 3:3, 4:4 and 5:5) are also present. Moreover, it was observed that the contribution of other resonant orbits (i.e., 2:1, 3:2, or higher) to the overall orbital structure of the system is extremely limited. In Fig. \ref{orbs}(a-h) we present an example of each of the basic types of the orbits. The box orbit and the chaotic orbit shown in Figs. \ref{orbs}a and \ref{orbs}h, respectively were computed until $t = 400$ time units, while all the parent\footnote{For every orbital family there is a parent (or mother) periodic orbit, that is, an orbit that describes a closed figure. Perturbing the initial conditions which define the exact position of a periodic orbit we generate quasi-periodic orbits that belong to the same orbital family and librate around their closed parent periodic orbit.} periodic orbits were computed until one period has completed. The black thick curve circumscribing each orbit in the physical $(x,y)$ plane is the Zero Velocity Curve (ZVC) which is defined as $V(x,y) = h$. In Table \ref{table} we provide the type, the initial conditions and the value of the energy for all the depicted orbits. In the resonant cases, the initial conditions and the period $T_{\rm per}$ correspond to the parent periodic orbits.

A very informative diagram the so-called ``characteristic" orbital diagram [\citealp{CM77}] is presented in Fig. \ref{otd}a. This diagram shows the evolution of the $y$ coordinate of the initial conditions of the parent periodic orbits of each orbital family as a function of their orbits energy $h$. Here we should emphasize, that for orbits starting perpendicular to the $y$-axis, we need only the initial condition of $y_0$ in order to locate them on this characteristic diagram. On the other hand, for orbits not starting perpendicular to the $y$-axis (i.e., the 1:1 linear and the 4:4 resonant orbits) initial conditions as position-velocity pairs $(y,\dot{y})$ are required and therefore, the characteristic diagram is now three-dimensional providing the full information regarding the interrelations of the initial conditions in a tree of families of periodic orbits. Furthermore, the diagram shown in Fig. \ref{otd}b is called the ``stability diagram" [\citealp{CB85}, \citealp{CM85}] and it illustrates the stability of all the families of periodic orbits in our dynamical system when the numerical value of the energy varies. A periodic orbit is stable if only the stability index (S.I.) [\citealp{MH92}, \citealp{Z13}] is between −2 and +2. This diagram help us monitor the evolution of S.I. of the resonant periodic orbits as well as the transitions from stability to instability and vice versa.

In the following Figs. \ref{grd1} and \ref{grd2} we present six grids of initial conditions $(x_0,y_0)$ and $(y_0, \dot{y_0})$ of orbits that we have classified for different values of the total orbital energy $h$ in the physical and in the phase space, respectively. The corresponding black thick curve circumscribing each grid in the phase $(y,\dot{y})$ plane is the limiting curve which can be obtained from
\begin{equation}
f(y,\dot{y}) = \frac{1}{2}\dot{y}^2 + V(x = 0, y) = h.
\label{limc}
\end{equation}
We should point out, that the grids in both the physical and the phase space have the form of a nut, being shaper in its corners when the value of energy is close to escape energy, while it gets smoother and more spherical as $h$ goes to zero. Looking at Figs. \ref{grd1} and \ref{grd2} it is evident that at very low energy levels the motion is entirely ordered. In particular, the vast majority of the regular orbits are either box or 1:1 resonant orbits which form three distinct stability islands. We also see that with increasing energy, the total area in the grids occupied by box orbits is constantly reduced and for $h > 0.06$ initial conditions corresponding to box orbits are mainly confined to the outer parts of the grids. Moreover, it is seen that when $h = 0.08$ a thin chaotic layer emerges inside the vast box region. For higher energy levels, the orbital structure of the computed grids alters significantly and the most noticeable changes are the following: (i) the initial conditions of box orbits practically vanish; (ii) the area occupied by 1:1 resonant orbits is reduced; (iii) the amount of chaotic orbits is highly increased and (iv) several bifurcated families of the main 1:1 family (i.e., the 2:2, 3:3, 4:4 and 5:5 resonant families) along with other higher resonant orbital families appear. Furthermore, we observe that for $h > 0.12$, that is when the value of the energy approaches very close to the escape energy, chaotic orbits is the most populated type of orbits however, small stability islands are still present and embedded inside the vast unified chaotic sea.

\begin{figure*}[!tH]
\centering
\resizebox{0.90\hsize}{!}{\includegraphics{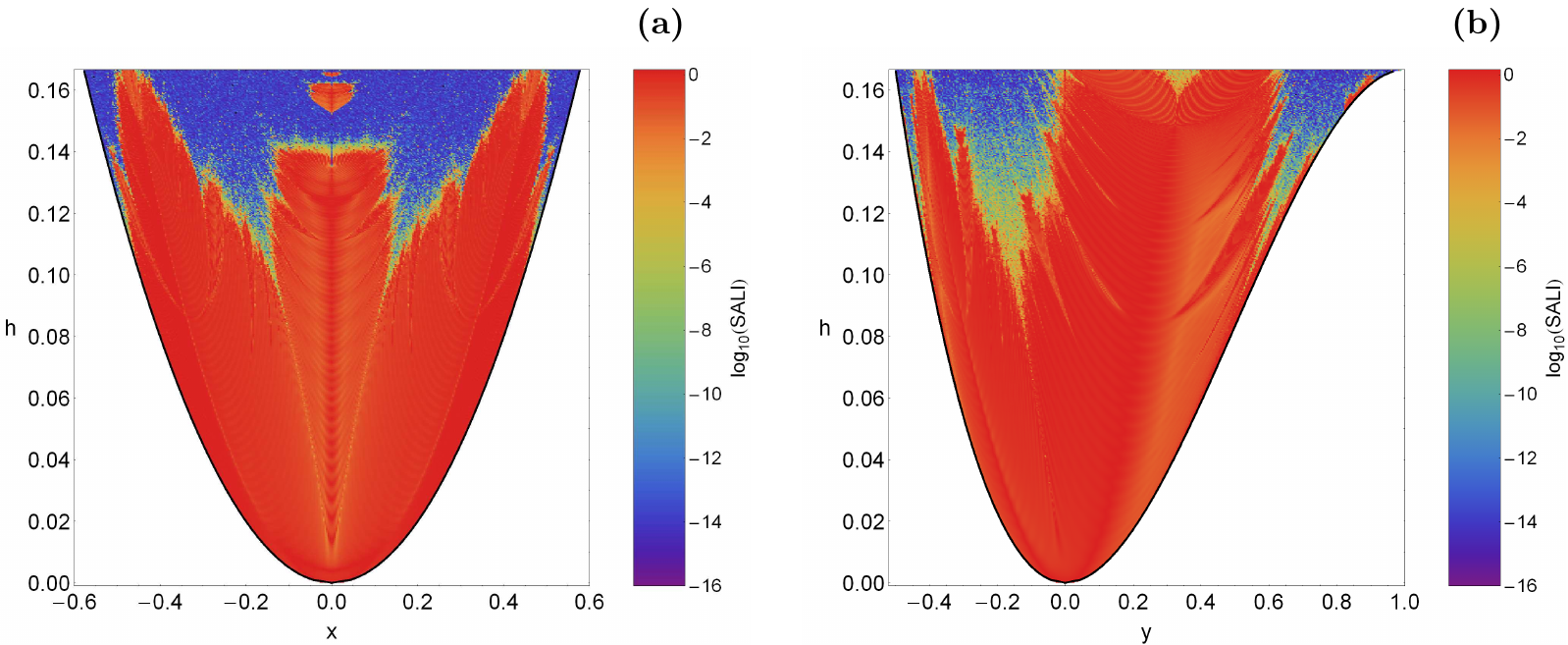}}
\caption{The orbital structure of the (a-left): $(x,h)$-plane and (b-right): $(y,h)$-plane. These diagrams give a detailed analysis of the evolution of the regular and chaotic orbits of the dynamical system when the value of the energy $h$ changes.}
\label{salis}
\end{figure*}

\begin{figure*}[!tH]
\centering
\resizebox{0.90\hsize}{!}{\includegraphics{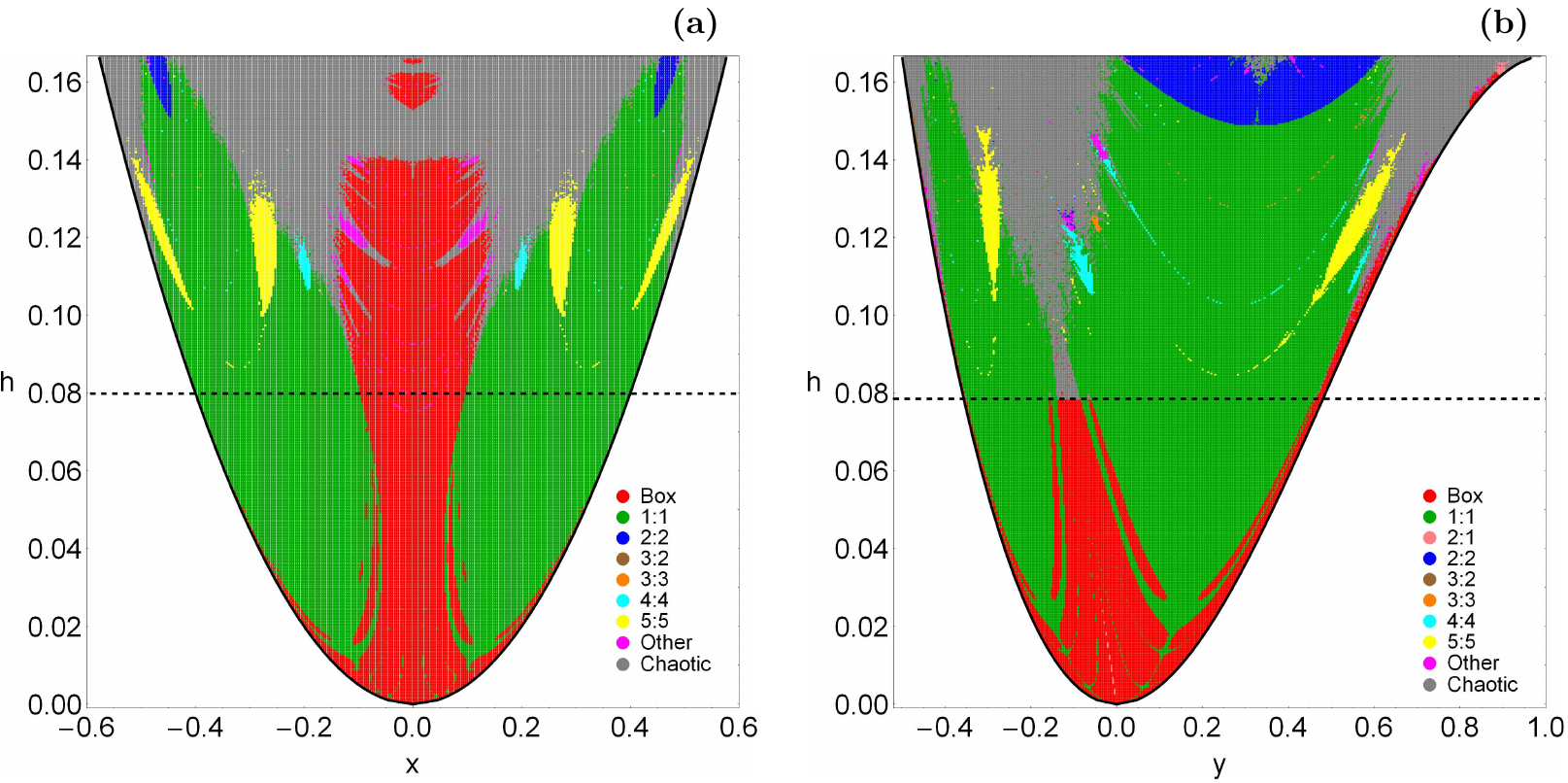}}
\caption{The orbital structure of the (a-left): $(x,h)$-plane and (b-right): $(y,h)$-plane. These diagrams give a detailed analysis of the evolution of the chaotic as well as of the different types of regular orbits of the dynamical system when the value of the energy $h$ changes.}
\label{clasif}
\end{figure*}

The evolution of the percentages of the chaotic orbits and that of the main families of regular orbits on the physical as well as on the phase space when the total orbital energy $h$ varies is shown in Fig. \ref{percs}(a-b). We observe, that the pattern of the curves is more or less the same in both the physical and the phase space. It is seen that for low values of the energy there is no evidence of chaotic motion. In particular, only two type of regular orbits are possible: box orbits which cover about 70\% of the total initial conditions and 1:1 resonant orbits which correspond to the remaining 30\% of the computed orbits. With increasing energy however, the percentage of the box drops rapidly, while at the same time the rate of the 1:1 resonant orbits grows significantly and very quickly $(h > 0.03)$ the 1:1 resonant orbits is the most populated family. In fact, in the interval $0.06 < h < 0.10$, the 1:1 resonant orbits occupy more than 90\% of the grids, while for larger values of the energy this tendency is reversed, their rate reduces rapidly and for $h = 0.16$ 1:1 resonant orbits occupy only about one tenth of the grids. The rate of box orbits on the other hand, is constantly being reduced and tends asymptotically to zero as we proceed to higher energy levels. Furthermore, we may say that chaos appears only at relatively high energies and when $h > 0.08$ the percentage of chaotic orbits exhibits a drastic increase and eventually chaotic orbits take over the grids when $h > 0.12$\footnote{It is known that in dynamical systems with a finite energy of escape where escape is possible (open Hamiltonian systems), chaos appears and dominates at values of energy very close to the critical escape energy which determines the transition from bounded to unbounded motion.}. The particular evolution of the percentage of chaotic orbits was anticipated because as the value of the energy increases we approach closer and closer to the critical value of the escape energy $(h_{esc} = 1/6)$. Indeed, at the higher energy value studied, that is when $h = 0.16$, about 90\% of the total initial conditions correspond to chaotic orbits. Moreover, we may argue that varying the value of the energy mainly shuffles the orbital content of all the other resonant orbits, whose percentages present fluctuations at low values (always less than 10\%). However, we would like to note that when $h = 0.12$, we observe a sudden spike at the percentages of 4:4 and 5:5 resonant orbits which belong to bifurcated families of the main 1:1 resonant family. In addition, the rate of the 2:2 resonant orbits is slightly elevated at $h = 0.16$ thus surpassing that of the main 1:1 family. Thus, taking into account all the above-mentioned analysis, we may conclude that in the H\'{e}non-Heiles system the total orbital energy $h$ influences mostly box, 1:1, and chaotic orbits. In fact, what really happens is that a large portion of 1:1 and box orbits turns into chaotic orbits as the value of the energy increases tending to the escape energy.

The grids in the physical $(x,y)$ as well as in the phase $(y,\dot{y})$ plane provide information on the phase space mixing for only a fixed value of energy. H\'{e}non however, back in the 60s, also considered two different types of planes which provide information about regions of regularity and chaos, using the value of the energy as an ordinate thus obtaining a continuous spectrum of values of $h$. The first type which is the $(x,h)$ plane uses the section $y = \dot{x} = 0$, $\dot{y} > 0$, i.e., the test particle starts on the $x$-axis, parallel to the $y$-axis and in the positive $y$-direction. Similarly, the second type, the $(y,h)$ plane uses the section $x = \dot{y} = 0$, $\dot{x} > 0$, i.e., the test particle starts on the $y$-axis, parallel to the $x$-axis and in the positive $x$-direction. In Fig. \ref{salis}(a-b) we show the structure of these planes when $h \in (0, h_{esc})$, where each initial condition is plotted in color code according to the value of SALI, thus distinguishing between ordered and chaotic motion. We observe that the vast majority of the computed orbits are regular, while on the other hand, initial conditions of chaotic orbits are mainly confined to right outer parts of the grids. In order to be able to monitor with sufficient accuracy and details the transition from regularity to chaos, we defined a dense grid of $10^5$ initial conditions in both cases. Here we should point out, that our results shown in Fig. \ref{salis}(a-b) coincide with that of Figs. 4 and 6, respectively presented in [\citealp{BBS08}], where the OFLI2 indicator was deployed in order to dissociate regular and chaotic orbits.

Finally, Fig. \ref{clasif}(a-b) presents another point of view of the planes discussed earlier in Fig. \ref{salis}(a-b). In this case, the initial conditions of the orbits are colored according to the regular or chaotic nature nature of the orbits giving emphasis to the different types of regular orbits. Once more, we observe that at low energy levels the vast majority of orbits are regular and specifically box orbits. As the value of the energy increases however, box orbits are depleted and 1:1 resonant orbits is the most populated family of regular orbits. It is also seen, that all the bifurcated families (i.e., the 2:2, 3:3, 4:4 and 5:5 families), as well as other higher resonant families of orbits (i.e., the 3:2 family) emerge at relatively high values of the energy (roughly about $h > 0.08$). Around the same energy level we also encounter the first initial conditions corresponding to chaotic motion. The horizontal black dashed line at Fig. \ref{clasif}(a-b) marks the first indication of chaos. Our numerical calculations suggest that at the $(x,h)$ plane chaotic motion appears at $h = 0.07916$, while at the $(y,h)$ plane the energy level in which chaos was detected for the first time equals to $h = 0.07882$.

\section{Discussion and conclusions}
\label{disc}

The aim of the present work was to investigate how influential is the total orbital energy on the level of chaos and on the distribution of the different regular families in the classical H\'{e}non-Heiles system. A very interesting feature of this Hamiltonian system is that it has a finite energy of escape. In particular, for energies smaller than the escape value, the equipotential surfaces are close and therefore escape is impossible. For energy levels larger than the escape energy on the other hand, the equipotential surfaces open and three channels of escape appear through which the particles can escape to infinity. However, it should be emphasized, that if a test particle has energy larger than the escape value, this does not necessarily mean that the particle will certainly escape from the system. In our research, we decided to consider only bounded motion and explore in details the regular or chaotic nature of orbits.

We numerically investigated the structure of both the physical $(x,y)$ and the phase $(y,\dot{y})$ space for a better understanding of the orbital properties of the system. For determining the regular or chaotic character of orbits in this system, we defined dense grids of initial conditions regularly distributed in the area allowed by the value of the total orbital energy $h$, in both the physical and the phase space. In both cases, the density of the grids was controlled in such a way that always there are at least 50000 orbits to be examined. For the numerical integration of the total set of orbits in each grid, we needed roughly between 17 to 40 hours of CPU time on a Pentium Dual-Core 2.2 GHz PC, depending mainly on the amount of regular orbits in each case which, unlike chaotic orbits, they required further numerical calculations in order to be classified into regular families.

To show how the value of the total energy $h$ influences the orbital structure of the system, we presented for each case, color-coded grids of initial conditions, which allow us to visualize what types of orbits occupy specific areas in the physical/phase space. Each orbit was integrated numerically for a time period of $10^4$ time units which corresponds to a time span of the order of hundreds of orbital periods. The particular choice of the total integration time was made in order to eliminate sticky orbits (classifying them correctly as chaotic orbits) with relatively long sticky periods. Then, we made a step further, in an attempt to distribute all regular orbits into different families. Therefore, once an orbit has been characterized as regular applying the SALI method, we then further classified it using a frequency analysis method. This method calculates the Fourier transform of the coordinates and velocities of an orbit, identifies its peaks, extracts the corresponding frequencies and then searches for the fundamental frequencies and their possible resonances.

Our thorough and detailed numerical analysis suggests that the level of chaos as well as the distribution in regular families are indeed very dependent on the total orbital energy. The main numerical results of our investigation can be summarized as follows:
\begin{enumerate}
 \item Several types of regular orbits were found to exist in H\'{e}non-Heiles potential, while there are also extended chaotic domains separating the areas of regularity. In particular, a large variety of resonant orbits are present, thus making the orbital structure of the system both rich and interesting.
 \item Our numerical computations revealed that most of the resonant families (i.e., the 2:2, 3:3, 4:4, 5:5 families) are in fact bifurcated families of the main 1:1 resonant family. Other types of resonant orbits on the other hand, (i.e., the 2:1, 3:2, etc) have extremely weak contribution to the overall orbital structure of the system.
 \item It was found that in both the physical $(x,y)$ and the phase $(y,\dot{y})$ space the total orbital energy influences mainly box, 1:1 resonant, and chaotic orbits. Moreover, for $h > 0.06$ the majority of test particles move either in 1:1 resonant or chaotic orbits.
 \item For low energy levels $(h < 0.08)$ the motion was found to be entirely regular, while for larger values of the energy the amount of chaotic orbits increases rapidly and for $h > 0.14$ they dominate occupying more than half of the entire phase space.
 \item The $(x,h)$ and $(y,h)$ diagrams gave us a more complete view of the evolution of the orbital families for a continuous spectrum of values of $h$. It was observed, that for low energy levels corresponding to local motion the orbital content is rather poor (only box and 1:1 resonant orbits are present), while most of the resonant orbits emerge at relatively high values of energy $(h > 0.08)$ suitable for global motion.
\end{enumerate}

We hope that the present analysis and the corresponding numerical results of the H\'{e}non-Heiles system to be useful in this active field of nonlinear Hamiltonian systems. Taking into account that our outcomes are encouraging, it is also in our future plans to expand our exploration in order to reveal the complete network of periodic orbits, thus shedding some light to the evolution of the periodic orbits as well as their stability when the orbital energy varies, considering also unbounded motion of particles.

\section*{Acknowledgments}

The author would like to express his warmest thanks to all the anonymous referees for the careful reading of the manuscript and for all the apt suggestions and comments which allowed us to improve both the quality and the clarity of the paper.

\end{document}